\documentclass[english]{llncs}
\usepackage[T1]{fontenc}
\usepackage[latin9]{inputenc}
\usepackage{verbatim}
\usepackage{amsmath}

\makeatletter
\usepackage[pdfpagescrop={92 62 523 728}]{hyperref}

\makeatother

\usepackage{babel}
\begin{document}

\title{Recognising and Generating Terms using Derivatives of Parsing Expression
Grammars}

\author{Tony Garnock-Jones,$^{1}$ Mahdi Eslamimehr,$^{2}$ and Alessandro Warth$\,{}^{2}$}

\institute{$^{1}$Northeastern University, Boston, Massachusetts, USA\\
$^{2}$Communications Design Group, Los Angeles, California, USA}
\maketitle
\begin{abstract}
Grammar-based sentence generation has been thoroughly explored for
Context-Free Grammars (CFGs), but remains unsolved for recognition-based
approaches such as Parsing Expression Grammars (PEGs). Lacking tool
support, language designers using PEGs have difficulty predicting
the behaviour of their parsers. In this paper, we extend the idea
of\emph{ derivatives}, originally formulated for regular expressions,
to PEGs. We then present a novel technique for sentence generation
based on derivatives, applicable to any grammatical formalism for
which the derivative can be defined\textemdash now including PEGs.
Finally, we propose applying derivatives more generally to other problems
facing language designers and implementers.
\end{abstract}

\section{\label{sec:Introduction}Introduction}

Writing a grammar that accurately captures the syntax of a new language
is challenging. No matter which formalism is used, seemingly correct
grammars often have bugs, accepting some inputs they should not and
rejecting others that they should. We would like to equip language
designers and implementers with tools that enable them to build a
precise understanding of the language that is accepted by a grammar.
Techniques for generating example sentences from a grammar can provide
a foundation for such a tool.

\emph{Parsing Expression Grammars} (PEGs) \cite{Ford2004} are recognition-based
formal descriptions of language syntax. They go beyond the expressiveness
of Context-Free Grammars (CFGs) by permitting unbounded lookahead,
both positive and negative, and offering a prioritized choice construct
corresponding to the recursive descent approach to alternation. These
features enable PEGs to recognise languages\textemdash such as $a^{n}b^{n}c^{n}$\textemdash that
CFGs cannot.

However, there is a price to pay: the interactions among these features
can be subtle. This often yields surprising results. A method for
generating example sentences following a PEG would help the programmer
understand its implications by providing concrete examples. For example,
in Ford's original paper on PEGs, he presents a grammar for the language
$a^{n}b^{n}c^{n}$, demonstrating the power of the formalism. It was
not until we used this grammar as a test case for our generation tool,
and it produced the sentence $aaa$, that we realised Ford's grammar
is not quite correct: it recognises sentences in $a^{+}$ in addition
to the intended $a^{n}b^{n}c^{n}$. Without the tool, it would not
have occurred to us to try such sentences with the grammar. We were
surprised by the flaw, showing that reading, implementing and using
the grammar was not enough to build sufficient understanding of it.
Note that this bug also eluded the inventor of the formalism, and
the POPL reviewers.\footnote{See Ford 2004 section 3.4 for his $a^{n}b^{n}c^{n}$ grammar \cite{Ford2004}.}

Generating sentences for CFGs has been thoroughly explored \cite{Godefroid2008,Majumdar2007,Dreyfus2013};
in fact, CFGs were originally proposed with sentence generation in
mind. PEGs, however, correspond directly to \emph{recursive descent}
parsing techniques \cite{Redziejowski2007}, and the potential for
sentence generation using PEGs has not been investigated until now.
Support for prioritized choice as well as positive and negative lookahead
in PEGs introduces constraints on sentence generation that make it
a more challenging task than the equivalent problem for CFGs. After
failing to approach the problem directly from a recursive-descent
perspective, we hit upon a technique based on the \emph{method of
derivatives}, used previously primarily for recognition \cite{Brzozowski1964,Owens2009,Might2011}.
The work presented in this paper builds on this idea, and makes the
following technical contributions:
\begin{itemize}
\item A definition of derivatives for PEGs.
\item A novel sentence generation technique based on derivatives that is
applicable to any grammatical formalism for which the derivative can
be defined.
\end{itemize}
Our general technique for sentence generation gives programmers much-needed
help in debugging their regular, context-free, and parsing expression
language specifications.

The rest of this paper is organized as follows. In section \ref{sec:Background},
we review the definition of PEGs and the method of recognising strings
using derivatives of grammars. Section \ref{sec:Generating-sentences}
describes our method of using derivatives for term generation. Section
\ref{sec:peg-derivatives} presents and justifies our method of computing
derivatives of PEGs. We touch on relevant implementation issues in
section \ref{sec:Implementing-derivatives}, and section \ref{sec:Evaluation}
argues for the correctness of our definitions. We discuss related
works in section \ref{sec:Related-Work}, outline the potential for
using derivatives in other ways to support language designers in section
\ref{sec:FutureWork}, and conclude the paper in section \ref{sec:Conclusion}.

\section{\label{sec:Background}Background}

\subsection{\label{subsec:Parsing-Expression-Grammars-definition}Parsing Expression
Grammars}

A PEG is defined as a tuple $G=(V_{N},V_{T},R,e_{S})$ of nonterminals
$V_{N}$, terminals $V_{T}$, rules $R:V_{N}\mapsto e$, and a start
expression $e_{S}$. Parsing expressions $e$ are defined inductively
as follows:
\begin{enumerate}
\item $\epsilon$, the empty string
\item $a$, any terminal, where $a\in V_{T}$
\item $A$, any nonterminal, where $A\in V_{N}$
\item $e_{1}\ e_{2},$ a sequence
\item $e_{1}/e_{2}$, an alternation
\item $e^{*},$ zero-or-more repetitions
\item $!e$, a not-predicate
\item $\_$, a wildcard
\item $\emptyset$, a \emph{failing} expression
\end{enumerate}
To simplify our presentation of derivatives, we extend the core PEG
expression variants (items 1 through 7 above) with the wildcard expression
$\_$ and the failing expression $\emptyset$ (items 8 and 9). We
omit positive lookahead $\&e$, since it can be defined as $\&e=!!e$
\cite[§3.2]{Ford2004}.

A parsing expression conditionally matches a \emph{prefix} of an input
sequence drawn from $V_{T}^{\star}$. We define the semantics of PEGs
with the function
\[
(e,x)\Rightarrow o
\]
where $x\in V_{T}^{*}$ and $o\in V_{T}^{*}\cup\{f\}$. (Note that
$f\notin V_{T}.)$ The ``output'' $o$ of a successful match is
the \emph{suffix }of\emph{ }the input stream that was not consumed
(recognized) by $e$, while an output of $f$ indicates failure. In
our definition of $\Rightarrow$ below, and in the remainder of the
paper, we write $a,b,c$ for terminals in $V_{T}$, $A,B,C$ for nonterminals
in $V_{N}$ and $x,y,z$ for (possibly-empty) strings in $V_{T}^{\star}$.
\begin{enumerate}
\item \textbf{Empty:} $(\epsilon,x)\Rightarrow x$
\item \textbf{Terminal (success case):} $(a,ax)\Rightarrow x$
\item \textbf{Terminal (failure case):} $(a,bx)\Rightarrow f$ if $a\neq b$
\item \textbf{Terminal (empty case):} $(a,\epsilon)\Rightarrow f$
\item \textbf{Nonterminal:} $(A,x)\Rightarrow o$ if $(R(A),x)\Rightarrow o$
\item \textbf{Sequence (progress case):} $(e_{1}\ e_{2},xy)\Rightarrow o$
if $(e_{1},xy)\Rightarrow y$ and $(e_{2},y)\Rightarrow o$
\item \textbf{Sequence (failure case):} $(e_{1}\ e_{2},x)\Rightarrow f$
if $(e_{1},x)\Rightarrow f$
\item \textbf{Alternation (success case): $(e_{1}/e_{2},xy)\Rightarrow y$
}if\textbf{ }$(e_{1},xy)\Rightarrow y$
\item \textbf{Alternation (fallback case): $(e_{1}/e_{2},x)\Rightarrow o$
}if\textbf{ }$(e_{1},x)\Rightarrow f$ and $(e_{2},x)\Rightarrow o$
\item \textbf{Zero-or-more repetitions (repetition case):} $(e^{*},xy)\Rightarrow o$
if $(e,xy)\Rightarrow y$ and $(e^{*},y)\Rightarrow o$
\item \textbf{Zero-or-more repetitions (termination case):} $(e^{*},x)\Rightarrow x$
if $(e,x)\Rightarrow f$
\item \textbf{Not-predicate (success case):} $(!e,x)\Rightarrow x$ if $(e,x)\Rightarrow f$
\item \textbf{Not-predicate (failure case):} $(!e,x)\Rightarrow f$ if $(e,x)\Rightarrow o$
and $o\neq f$
\item \textbf{Wildcard (success case):} $(\_,ax)\Rightarrow x$
\item \textbf{Wildcard (empty case):} \textbf{$(\_,\epsilon)\Rightarrow f$}
\item \textbf{Failing:} $(\emptyset,x)\Rightarrow f$
\end{enumerate}
Note that the expression $e_{1}/e_{2}$ denotes a \emph{prioritized}
choice: it is only when $e_{1}$ fails to match a given input that
$e_{2}$ is given a chance. This is a key feature of PEGs, making
them unambiguous, but is also one of the chief difficulties in reasoning
about PEG behaviour while developing a grammar.

\subsection{\label{subsec:Recognising-with-derivatives}Recognising with derivatives}

The method of derivatives, introduced for regular expressions (REs)
in 1964 by Brzozowski \cite{Brzozowski1964}, revisited in 2009 by
Owens et al. \cite{Owens2009}, and extended to CFGs in 2011 by Might
et al. \cite{Might2011}, is a powerful, easily-understood and easily-implemented
recognising technique.

The key idea is to define a \emph{derivative }function that maps a
grammar $G$ and a terminal $a$ to a derived grammar that should
match the ``rest of the input'' after $a$ has been consumed. Deciding
whether a given input sequence is in the language of $G$ is done
by iterating over successive input tokens until the end of input is
reached. At that point, the final derived grammar is examined. If
its language includes the empty string, then the input is in the language
of $G$; otherwise, it is not.

Formally, the derivative function $D$ has type $V_{T}\times e\rightarrow e$.
For REs, $V_{T}$ is the input alphabet, and $e$ is a regular expression;
for CFGs, $e$ is extended to permit \emph{recursion} \cite{Might2011}
and an environment mapping nonterminals to expressions is assumed;
and for PEGs, $V_{T}$ and $e$ are to be understood as relating to
a particular grammar $G$ as defined in section \ref{subsec:Parsing-Expression-Grammars-definition}.

We write $D_{a}e$ (the derivative of $e$ with respect to $a$) for
$a\in V_{T}$, and extend $D$ inductively to sequences $x\in V_{T}^{\star}$
by
\begin{eqnarray*}
D_{\epsilon}e & = & e\\
D_{ax}e & = & D_{x}(D_{a}e)
\end{eqnarray*}

In order to recognise using derivatives, we must be able to reliably
determine whether the empty sequence $\epsilon$ is in the language
of a particular expression, $L(e)$. To do this, we use a \emph{nullability
predicate},\footnote{Nullability is defined for REs by Brzozowski \cite{Brzozowski1964}
and for CFGs by Might et al. \cite{Might2011}} 
\[
\nu(e)\iff\epsilon\in L(e)
\]

Deciding whether a non-empty sequence $ax$ is in $L(e)$ is done
by deciding whether $x$ is in $L(D_{a}e)$. Summing up, and using
the extension of $D$ to sequences given above,
\[
x\in L(e)\iff\nu(D_{x}e)
\]

Both $\nu(e)$ and $D_{a}e$ are defined for PEGs below. The definition
of $\nu(e)$ for PEGs includes an additional constraint not relevant
to REs or CFGs: that, at the point of the nullability check, no part
of $e$ needs to perform any further lookahead in order to accept.

\section{\label{sec:Generating-sentences}Generating sentences using derivatives}

\begin{figure}[tb]
\begin{eqnarray*}
\mathit{gen} & : & e\rightarrow V_{T}^{\star}\cup\{f\}\\
\mathit{gen}\;e & = & \begin{cases}
\mathit{gen'}\;e\;\emptyset & \textrm{when enough output has been generated}\\
\mathit{gen'}\;e\;(\mathit{firsts}\;e) & \textrm{otherwise}
\end{cases}\\
\\
\mathit{gen'} & : & e\times\mathcal{P}(V_{T})\rightarrow V_{T}^{\star}\cup\{f\}\\
\mathit{gen'}\;e\;(\{a\}\uplus t) & = & \begin{cases}
ax & \mathrm{when}\;\mathit{gen}\;D_{a}e=x\\
\mathit{gen'}\;e\;t & \mathrm{when}\;\mathit{gen}\;D_{a}e=f
\end{cases}\\
\mathit{gen'}\;e\;\emptyset & = & \begin{cases}
\epsilon & \mathrm{when}\;\nu(e)\\
f & \mathrm{otherwise}
\end{cases}
\end{eqnarray*}

\caption{\label{fig:generation-algorithm}Generic algorithm for generating
sentences using derivative, nullability, and first sets.}
\end{figure}
\begin{figure}
\begin{eqnarray*}
\mathit{firsts} & : & e\rightarrow\mathcal{P}(V_{T})\\
\mathit{firsts}\;e & = & \mathit{firsts'}\;e\;\emptyset\\
\\
\mathit{firsts'} & : & e\times\mathcal{P}(V_{T})\rightarrow\mathcal{P}(V_{T})\\
\mathit{firsts'}\;\epsilon\;t & = & t\\
\mathit{firsts'}\;a\;t & = & \{a\}\\
\mathit{firsts'}\;A\;t & = & \mathit{firsts'}\;R(A)\ t\\
\mathit{firsts'}\;e_{1}\ e_{2}\;t & = & \mathit{firsts'}\;e_{1}\;(\mathit{firsts'}\;e_{2}\;t)\\
\mathit{firsts'}\;e_{1}/e_{2}\;t & = & \mathit{firsts'}\;e_{1}\;t\cup\mathit{firsts'}\;e_{2}\;t\\
\mathit{firsts'}\;e^{\star}\;t & = & t\cup\mathit{firsts'}\;e\;t\\
\mathit{firsts'}\;!e\;t & = & t\\
\mathit{firsts'}\;\_\;t & = & V_{T}\\
\mathit{firsts'}\;\emptyset\;t & = & \emptyset
\end{eqnarray*}

\caption{\label{fig:peg-firsts-algorithm}Algorithm for (over)approximating
the first set of a PEG.}

\end{figure}
Equipped with a definition of the derivative of a grammar, we can
now turn to its application in generating sentences within that grammar.
Our generation method, the function $\mathit{gen}$ in figure \ref{fig:generation-algorithm},
is generic in the sense that it is valid for any grammatical formalism
for which a nullability predicate and derivative function can be defined.

The only additional requirement is that some overapproximation to
the \emph{first set} \cite{Aho1986} of a grammar can be computed.
Any definition of the function $\mathit{firsts}$ will work, so long
as it really is an overapproximation of the actual first set of its
argument. Even setting $\mathit{firsts}\;e=V_{T}$ would generate
correct output, since $\mathit{gen'}$ explores other alternatives
if a selected terminal leads to failure of generation.

The algorithm $\mathit{gen}$ operates by selecting some terminal
$a$ from the first set of a parsing expression $e$, and then recursing
with the corresponding derivative $D_{a}e$, prepending $a$ to the
result. The algorithm can continue to randomly produce terminals as
long as elements from the first set of $e$ at each stage remain unexplored,
and may yield the empty output sequence any time that $e$ is nullable.
Any criterion for placing an upper bound on the length of the output
sequence may be used.

Our definition of $\mathit{firsts}$ for PEGs (figure \ref{fig:peg-firsts-algorithm})
avoids much inefficiency by overapproximating only in the case of
negative lookahead, $!e$. Consider the expression $!(abc)\;(a/b/c)^{\star}$.
The first set of that expression must include $a$, even though $a$
is in the first set of the expression $abc$, because $acb$ is accepted
by the overall expression, even though $abc$ is rejected. In general,
we cannot use $\mathit{firsts}\;e$ when computing $\mathit{firsts'}\;!e\;t$
because $e$ may examine more than a single token's worth of its input.

Because we overapproximate the first set of $!e$, and we treat \emph{positive}
lookahead $\&e$ as if it were $!!e$, we overapproximate the first
set of $\&e$ as well. A practical implementation may choose to include
$\&e$ as first-class syntax, defining $\mathit{firsts'}\;\&e\;t=\mathit{firsts}\;e\cap t$
in order to avoid unnecessary work in $\mathit{gen}$.

\section{\label{sec:peg-derivatives}Derivatives of PEGs}

\begin{figure}[tb]
\begin{minipage}[t]{0.49\columnwidth}%
\begin{eqnarray*}
D & : & V_{T}\times e\rightarrow e\\
D_{a}\epsilon & = & \emptyset\\
D_{a}a & = & \epsilon\\
D_{a}b & = & \emptyset\;\mathrm{when}\;a\not=b\\
D_{a}A & = & D_{a}R(A)\\
D_{a}(e_{1}\ e_{2}) & = & D_{a}e_{1}\;e_{2}\ /\ \delta_{a}e_{1}\;D_{a}e_{2}\\
D_{a}(e_{1}/e_{2}) & = & D_{a}e_{1}\ /\ D_{a}e_{2}\\
D_{a}e^{\star} & = & D_{a}e\;e^{\star}\;/\;\delta_{a}e\;D_{a}e^{\star}\\
D_{a}!e & = & \emptyset\\
D_{a}\_ & = & \epsilon\\
D_{a}\emptyset & = & \emptyset
\end{eqnarray*}
\end{minipage}%
\begin{minipage}[t]{0.49\columnwidth}%
\begin{eqnarray*}
\delta & : & V_{T}\times e\rightarrow e\\
\delta_{a}\epsilon & = & \epsilon\\
\delta_{a}b & = & \emptyset\\
\\
\delta_{a}A & = & \delta_{a}R(A)\\
\delta_{a}(e_{1}\ e_{2}) & = & \delta_{a}e_{1}\ \delta_{a}e_{2}\\
\delta_{a}(e_{1}/e_{2}) & = & \delta_{a}e_{1}\ /\ \delta_{a}!e_{1}\ \delta_{a}e_{2}\\
\delta_{a}e^{\star} & = & \delta_{a}e\;\delta_{a}e^{\star}\;/\;\delta_{a}!e\;\delta_{a}e^{\star}\\
\delta_{a}!e & = & !D_{a}(e\quad\_^{*})\\
\delta_{a}\_ & = & \emptyset\\
\delta_{a}\emptyset & = & \emptyset
\end{eqnarray*}
\end{minipage}

\caption{\label{fig:derivative-and-delta-definition}Derivative function $D$
and the nullability combinator $\delta$ for PEGs. }
\end{figure}
Our definition of the derivative function for PEGs (figure \ref{fig:derivative-and-delta-definition})
is closely modelled on Brzozowski's original definition for REs. The
chief differences arise from the complications of \emph{lookahead},
both positive and negative, and \emph{prioritized choice}.

\subsection{Lookahead}

In most implementations of PEGs, the expression $\&e_{1}\;e_{2}$
is evaluated by matching the input against $e_{1}$ and then, if the
match succeeds, rewinding the input and matching it with $e_{2}$.
Negative lookahead $!e_{1}\;e_{2}$ is evaluated similarly, rewinding
and turning to $e_{2}$ only when matching against $e_{1}$ fails.
This sequential processing is reflected in the use of the sequencing
operator to compose a lookahead with another expression.

Derivatives, by contrast, have no notion of backtracking, rewinding,
or sequential processing. Each input token is presented to the derivative
function only once, and all possible alternative matches proceed essentially
in \emph{parallel}. The lookahead problem, then, becomes a special
case of the awkward problem of sequencing in general.

Our treatment of sequencing follows the established pattern of using
a nullability \emph{combinator}, $\delta$, to properly account for
the non-com\-mut\-at\-iv\-ity of the sequencing operator. Brzozowski's
rule for the derivative of the concatenation of two regular expressions
$R_{1}$ and $R_{2}$ is
\[
D_{a}(R_{1}\;R_{2})=(D_{a}R_{1})\;R_{2}\;+\;\delta R_{1}\;(D_{a}R_{2})
\]

The term $\delta R_{1}$ evaluates to $\epsilon$ if $R_{1}$ is nullable,
and to $\emptyset$ otherwise. The intuition behind this rule is that
if any string in $L(R_{1})$ begins with $a$, then $D_{a}R_{1}\not=\emptyset$
and so $D_{a}(R_{1}R_{2})$\textemdash the grammar for the ``rest''
of the input\textemdash must include all strings beginning with the
``rest'' of $R_{1}$ and continuing with the strings of $L(R_{2})$.
In addition, if $L(R_{1})$ includes the empty string ($\delta R_{1}=\epsilon$),
then we must examine the strings in $L(R_{2})$ to see if any begin
with $a$. Otherwise, if $L(R_{1})$ does not include the empty string
($\delta R_{1}=\emptyset$), then there is some unmatched requirement
of $R_{1}$ that should prevent us from examining the strings in $L(R_{2})$.
In this case, the right-hand branch of the sum is effectively discarded.

This intuition extends to PEGs with minor modification for proper
treatment of lookahead. In general, PEG sequences will be of the form
\[
\&e_{1}\;...\;\&e_{n}\;!e_{n+1}\;...\;!e_{m}\;e_{0}
\]
where we effectively expect each $e_{i}$'s processing of the input
string to proceed in parallel. The PEG version of the nullability
combinator (figure \ref{fig:derivative-and-delta-definition}) must
not only check whether $\epsilon\in L(e_{0})$, but \emph{in the case
that it is}, must also advance each of $e_{1}...e_{m}$, since their
processing of the input should not be limited by the behaviour of
$e_{0}$. That is, we must ensure that
\[
D_{a}(\&e_{1}\;...\;\&e_{n}\;!e_{n+1}\;...\;!e_{m}\;e_{0})=\&D_{a}e_{1}\;...\;\&D_{a}e_{n}\;!D_{a}e_{n+1}\;...\;!D_{a}e_{m}\;D_{a}e_{0}
\]
since \emph{all} of the $e_{i>0}$ must examine $a$ in parallel with
$e_{0}$.

Our nullability combinator therefore takes an additional argument,
$a$, so that it can properly pass it on to $D$ in cases involving
lookahead. Our rule for computing the derivative of a sequence expression
becomes
\[
D_{a}(e_{1}\ e_{2})=(D_{a}e_{1})\;e_{2}\ /\ \delta_{a}e_{1}\;(D_{a}e_{2})
\]
and in the specific case of $!e_{1}\;e_{2}$,
\begin{eqnarray}
D_{a}(!e_{1}\;e_{2}) & = & (D_{a}!e_{1})\;e_{2}\;/\;\delta_{a}!e_{1}\;(D_{a}e_{2})\nonumber \\
 & = & \emptyset\;e_{2}\;/\;!(D_{a}e_{1})\;(D_{a}e_{2})\nonumber \\
 & = & \emptyset\;/\;!(D_{a}e_{1})\;(D_{a}e_{2})\nonumber \\
 & = & !(D_{a}e_{1})\;(D_{a}e_{2})\label{eq:seq-of-neg-lookahead}
\end{eqnarray}
as desired.\footnote{While working through examples such as this, we will regularly be
simplifying as we go using the identities discussed in section \ref{sec:Implementing-derivatives}.} The term $\delta_{a}!e_{1}$ evaluates to $!(D_{a}e_{1})$ on the
right-hand side of the alternation. The left-hand side of the alternation
is discarded entirely, because $D_{a}!e_{1}=\emptyset$ for any $e_{1}$.
\begin{example}
Consider the PEG
\[
P\ \leftarrow\ \&abc\quad\_\,\_\,\_
\]
After the input $a$, the ``rest'' of the grammar is $D_{a}P$:
\begin{eqnarray*}
D_{a}P & = & D_{a}(\&abc\quad\_\,\_\,\_)\\
 & = & D_{a}(\&abc)\quad\_\,\_\,\_\quad/\quad\delta_{a}(\&abc)\quad D_{a}(\_\,\_\,\_)\\
 & = & \emptyset\quad\_\,\_\,\_\quad/\quad\&(D_{a}(abc\quad\_^{*}))\quad D_{a}(\_\,\_\,\_)\\
 & = & \emptyset\quad/\quad\&(bc\quad\_^{*})\quad\_\,\_\\
 & = & \&(bc\quad\_^{*})\quad\_\,\_\\
 & = & \&bc\quad\_\,\_
\end{eqnarray*}
Notice that the left-hand branch of the alternation is always pruned
when computing the derivative of a sequence where the first element
is a lookahead. If it were not so, the derived grammar would \emph{only}
present the lookahead with the next input character, and other branches
of the multiple parallel ongoing parses would be incorrectly stalled.
Notice also that $\delta$ ``steps'' the lookahead in the right-hand
branch of the alternation in parallel with the ``stepping'' of the
remainder of the grammar.
\end{example}

\begin{property}
Consider now the general case of $(!e_{1}\ e_{2})\ e_{3}$. Computing
the derivative must respect the associativity of sequencing, i.e.,
\[
D_{a}((!e_{1}\;e_{2})\;e_{3})=D_{a}(!e_{1}\ (e_{2}\ e_{3}))
\]
\end{property}

\begin{proof}
\begin{eqnarray*}
D_{a}((!e_{1}\ e_{2})\ e_{3}) & = & D_{a}(!e_{1}\;e_{2})\;e_{3}\;/\;\delta_{a}(!e_{1}\;e_{2})\;D_{a}e_{3}\\
 & = & !(D_{a}e_{1})\;(D_{a}e_{2})\;e_{3}\;/\;\delta_{a}(!e_{1}\;e_{2})\;D_{a}e_{3}\quad\textrm{(by (\ref{eq:seq-of-neg-lookahead}))}\\
 & = & !(D_{a}e_{1})\;(D_{a}e_{2})\;e_{3}\;/\;\delta_{a}!e_{1}\;\delta_{a}e_{2}\;D_{a}e_{3}\\
 & = & !(D_{a}e_{1})\;(D_{a}e_{2})\;e_{3}\;/\;!(D_{a}e_{1})\;\delta_{a}e_{2}\;D_{a}e_{3}\\
 & = & !(D_{a}e_{1})\;((D_{a}e_{2})\;e_{3}\;/\;\delta_{a}e_{2}\;D_{a}e_{3})\quad\textrm{(by left-factoring)}\\
 & = & !(D_{a}e_{1})\;D_{a}(e_{2}\;e_{3})\\
 & = & D_{a}(!e_{1}\;(e_{2}\;e_{3}))
\end{eqnarray*}
\end{proof}

\subsection{Prioritized choice}

Lookahead is not the only complication in defining the derivatives
of PEGs. Instead of the unbiased sum operator of CFGs, PEGs make use
of a \emph{prioritized choice} operator. When evaluating a choice
$e_{1}/e_{2}$, if $e_{1}$ accepts (a prefix of) the input, then
$e_{2}$ is not even visited by the traditional PEG algorithm. Our
definition of $D_{a}(e_{1}/e_{2})$ must reflect this bias. We approach
the problem by observing that
\begin{equation}
e_{1}/e_{2}\equiv e_{1}/!(e_{1})e_{2},\label{eq:prioritized-equivalence}
\end{equation}
which makes clear that any inputs accepted by $e_{1}$ prevent the
right-hand branch of the alternation from accepting. Further, if alternation
in PEGs \emph{were} unbiased, then both $D_{a}$ and $\delta_{a}$
would simply distribute across alternation, just as they do for alternation
in REs and CFGs. Hypothetically, then:
\begin{eqnarray}
D_{a}(e_{1}/e_{2}) & \overset{?}{=} & D_{a}e_{1}/D_{a}e_{2}\nonumber \\
\delta_{a}(e_{1}/e_{2}) & \overset{?}{=} & \delta_{a}e_{1}/\delta_{a}e_{2}\label{eq:hypothetical-delta-alt}
\end{eqnarray}

These definitions make a reasonable starting point in our search for
a means of accommodating prioritized choice. In the case of $D_{a}$,
the intuition is that the input terminal $a$ could relate equally
to both possible alternatives. In the case of $\delta_{a}$, which
is used only in our rule for sequence expressions, we must evaluate
to a failing expression only if \emph{both} $\delta_{a}e_{1}$ and
$\delta_{a}e_{2}$ evaluate to a failing expression; otherwise, we
might prune the derived grammar inappropriately.

Applying (\ref{eq:prioritized-equivalence}) to these starting points,
we learn that the bias in PEG alternation does not affect the definition
of $D$: 
\begin{eqnarray*}
D_{a}(e_{1}/e_{2}) & \equiv & D_{a}(e_{1}/!(e_{1})e_{2})\textrm{ (by (\ref{eq:prioritized-equivalence}))}\\
 & \overset{?}{=} & D_{a}e_{1}/D_{a}(!(e_{1})e_{2})\\
 & = & D_{a}e_{1}/(D_{a}!e_{1}\;e_{2}\;/\;\delta_{a}!e_{1}\;D_{a}e_{2})\\
 & = & D_{a}e_{1}/(\emptyset\;e_{2}\;/\;!D_{a}e_{1}\;D_{a}e_{2})\\
 & = & D_{a}e_{1}/!D_{a}e_{1}\;D_{a}e_{2}\\
 & \equiv & D_{a}e_{1}/D_{a}e_{2}\textrm{ (by (\ref{eq:prioritized-equivalence}))}
\end{eqnarray*}
but that the definition of $\delta$ must take the bias into account:
\begin{eqnarray*}
\delta_{a}(e_{1}/e_{2}) & \equiv & \delta_{a}(e_{1}/!(e_{1})e_{2})\\
 & \overset{?}{=} & \delta_{a}e_{1}/\delta_{a}(!(e_{1})e_{2})\\
 & = & \delta_{a}e_{1}/\delta_{a}!e_{1}\;\delta_{a}e_{2}\\
 & = & \delta_{a}e_{1}/!D_{a}e_{1}\;\delta_{a}e_{2}\\
 & \not\equiv & \delta_{a}e_{1}/\delta_{a}e_{2}
\end{eqnarray*}
We therefore define
\[
\delta_{a}(e_{1}/e_{2})=\delta_{a}e_{1}\;/\;\delta_{a}!e_{1}\;\delta_{a}e_{2}
\]
not only to follow our intuition, outlined above, but also to ensure
that the branches are \emph{disjoint} \cite[§3.7]{Ford2004} so that
$e_{2}$ only influences the result when $e_{1}$ has certainly failed.
\begin{example}
Consider the PEG
\[
Q\leftarrow\&((a/\epsilon)\;!b)\;\_^{\star}
\]

If we attempt to compute $D_{a}Q$ using our hypothetical definition
(\ref{eq:hypothetical-delta-alt}), we get:
\begin{eqnarray}
D_{a}Q & \leftarrow & D_{a}(\&((a/\epsilon)\;!b)\;\_^{\star})\nonumber \\
 &  & \textrm{\{apply sequence rule for \ensuremath{D}\}}\nonumber \\
 & = & D_{a}\&((a/\epsilon)\;!b)\;\_^{\star}\;/\;(\delta_{a}\&((a/\epsilon)\;!b))\;D_{a}(\_^{\star})\nonumber \\
 &  & \textrm{\{apply \ensuremath{D_{a}\&e=\emptyset}\}}\nonumber \\
 & = & \emptyset\;\_^{\star}\;/\;(\delta_{a}\&((a/\epsilon)\;!b))\;D_{a}(\_^{\star})\nonumber \\
 &  & \textrm{\{apply \ensuremath{\delta_{a}\&e=\&D_{a}(e\;\_^{\star})} and \ensuremath{\emptyset\;e=\emptyset} and \ensuremath{D_{a}\_^{\star}=\_^{\star}}\}}\nonumber \\
 & = & \emptyset\;/\;\&(D_{a}(((a/\epsilon)\;!b)\;\_^{\star}))\;\_^{\star}\nonumber \\
 &  & \textrm{\{apply sequence rule for \ensuremath{D}\}}\nonumber \\
 & = & \emptyset\;/\;\&(D_{a}((a/\epsilon)\;!b)\;\_^{\star}\;/\;\delta_{a}((a/\epsilon)\;!b)\;D_{a}(\_^{\star}))\;\_^{\star}\nonumber \\
 &  & \textrm{\{apply sequence rule for \ensuremath{D} and \ensuremath{\emptyset/e=e}\}}\nonumber \\
 & = & \&((D_{a}(a/\epsilon)\;!b\;/\;\delta_{a}(a/\epsilon)\;D_{a}!b)\;\_^{\star}\;/\;\delta_{a}((a/\epsilon)\;!b)\;D_{a}(\_^{\star}))\;\_^{\star}\nonumber \\
 &  & \textrm{\{apply choice rule for \ensuremath{D} and sequence rule for \ensuremath{\delta} and \ensuremath{D_{a}\_^{\star}=\_^{\star}}\}}\nonumber \\
 & = & \&(((D_{a}a/D_{a}\epsilon)\;!b\;/\;\delta_{a}(a/\epsilon)\;D_{a}!b)\;\_^{\star}\;/\;(\delta_{a}(a/\epsilon)\;\delta_{a}!b)\;\_^{\star})\;\_^{\star}\nonumber \\
 &  & \textrm{\{apply base cases for \ensuremath{D}\}}\nonumber \\
 & = & \&(((\epsilon/\emptyset)\;!b\;/\;\delta_{a}(a/\epsilon)\;\emptyset)\;\_^{\star}\;/\;(\delta_{a}(a/\epsilon)\;\delta_{a}!b)\;\_^{\star})\;\_^{\star}\nonumber \\
 &  & \textrm{\{simplify with \ensuremath{e/\emptyset=e}, \ensuremath{\epsilon\;e=e}, \ensuremath{e\;\emptyset=\emptyset}\}}\nonumber \\
 & = & \&(!b\;\_^{\star}\;/\;(\delta_{a}(a/\epsilon)\;\delta_{a}!b)\;\_^{\star})\;\_^{\star}\nonumber \\
 &  & \textrm{\{apply \ensuremath{\delta_{a}!b=!D_{a}(b\;\_^{\star})=\dots=!(\emptyset\;\_^{\star})=!\emptyset=\epsilon}\}}\nonumber \\
 & = & \&(!b\;\_^{\star}\;/\;(\delta_{a}(a/\epsilon)\;\epsilon)\;\_^{\star})\;\_^{\star}\label{eq:point-of-difference-in-example}\\
 &  & \textrm{\{apply equation (\ref{eq:hypothetical-delta-alt})\}}\nonumber \\
 & = & \&(!b\;\_^{\star}\;/\;((\delta_{a}a/\delta_{a}\epsilon)\;\epsilon)\;\_^{\star})\;\_^{\star}\nonumber \\
 &  & \textrm{\{apply base cases for \ensuremath{\delta}\}}\nonumber \\
 & = & \&(!b\;\_^{\star}\;/\;((\emptyset/\epsilon)\;\epsilon)\;\_^{\star})\;\_^{\star}\nonumber \\
 &  & \textrm{\{simplify with \ensuremath{\emptyset/\epsilon=\epsilon} and \ensuremath{\epsilon\;\epsilon=\epsilon}\}}\nonumber \\
 & = & \&(!b\;\_^{\star}\;/\;\epsilon\;\_^{\star})\;\_^{\star}\nonumber \\
 &  & \textrm{\{simplify with \ensuremath{e_{1}\_^{\star}/e_{2}\_^{\star}=(e_{1}/e_{2})\_^{\star}} and \ensuremath{\&(e\;\_^{\star})=\&e}\}}\nonumber \\
 & = & \&(!b\;/\;\epsilon)\;\_^{\star}\nonumber 
\end{eqnarray}
This is incorrect, however. Since $!b/\epsilon$ is equivalent to
$\epsilon$, the faulty $D_{a}Q$ above is equivalent to $\_^{\star}$.
The constraint preventing a $b$ terminal from appearing next has
been lost.

If instead we use $\delta$ as defined in figure \ref{fig:derivative-and-delta-definition}
to compute $D_{a}Q$, then repeating our calculation from the point
of difference (marked (\ref{eq:point-of-difference-in-example}) above),
we see that
\begin{eqnarray*}
D_{a}Q & \leftarrow & D_{a}(\&((a/\epsilon)\;!b)\;\_^{\star})\\
 &  & \dots\\
 & = & \&(!b\;\_^{\star}\;/\;(\delta_{a}(a/\epsilon)\;\epsilon)\;\_^{\star})\;\_^{\star}\\
 &  & \textrm{\{apply correct choice rule for \ensuremath{\delta} from figure \ref{fig:derivative-and-delta-definition}\}}\\
 & = & \&(!b\;\_^{\star}\;/\;((\delta_{a}a\;/\;\delta_{a}!a\;\delta_{a}\epsilon)\;\epsilon)\;\_^{\star})\;\_^{\star}\\
 &  & \textrm{\{apply base cases for \ensuremath{\delta}\}}\\
 & = & \&(!b\;\_^{\star}\;/\;((\emptyset\;/\;\delta_{a}!a\;\epsilon)\;\epsilon)\;\_^{\star})\;\_^{\star}\\
 &  & \textrm{\{apply \ensuremath{\delta_{a}!a=!D_{a}(a\;\_^{\star})=\dots=!(\epsilon\;\_^{\star})=!(\_^{\star})=\emptyset}\}}\\
 & = & \&(!b\;\_^{\star}\;/\;((\emptyset\;/\;\emptyset\;\epsilon)\;\epsilon)\;\_^{\star})\;\_^{\star}\\
 &  & \textrm{\{simplify with \ensuremath{\emptyset/e=e} and \ensuremath{\emptyset\ e=\emptyset}\}}\\
 & = & \&(!b\;\_^{\star}\;/\;\emptyset)\;\_^{\star}\\
 &  & \textrm{\{simplify with \ensuremath{e/\emptyset=e} and \ensuremath{\&(e\;\_^{\star})=\&e} and \ensuremath{\&!e=!e}\}}\\
 & = & !b\;\_^{\star}
\end{eqnarray*}
which correctly rejects the input if the next token is $b$.
\end{example}

\subsection{Nullability and Well-formedness}

The algorithms for recognising and for generating sentences using
derivatives (sections \ref{subsec:Recognising-with-derivatives} and
\ref{sec:Generating-sentences}, respectively) need the nullability
predicate $\nu$ to decide when to stop. A straightforward extension
of Brzozowski's definition of $\nu$ suffices for \emph{well-formed}
PEGs, i.e., grammars that do not contain left-recursive rules \cite{Ford2004}.
Figure \ref{fig:Nullability-and-WF} defines the nullability and well-formedness
(WF) predicates for PEGs.

For a well-formed PEG $e$, nullability $\nu(e)$ means that $(e,\epsilon)\Rightarrow\epsilon$,
and vice versa. In other words, a PEG is considered nullable when,
given the empty input, the decision can be made to accept immediately,
without need to examine further input to decide questions posed by
use of lookahead. In Ford's terms, this is equivalent to a successful
parse when presented with the empty input sequence.

\begin{figure}[tb]
\begin{minipage}[t]{0.49\columnwidth}%
\begin{eqnarray*}
\nu & : & e\rightarrow2\\
\nu(\epsilon) & = & \top\\
\nu(a) & = & \bot\\
\nu(A) & = & \nu(R(A))\\
\nu(e_{1}e_{2}) & = & \nu(e_{1})\wedge\nu(e_{2})\\
\nu(e_{1}/e_{2}) & = & \nu(e_{1})\vee\nu(e_{2})\\
\nu(e^{\star}) & = & \top\\
\nu(!e) & = & \neg\nu(e)\\
\nu(\_) & = & \bot\\
\nu(\emptyset) & = & \bot
\end{eqnarray*}
\end{minipage}%
\begin{minipage}[t]{0.49\columnwidth}%
\begin{eqnarray*}
\mathit{WF} & : & e\rightarrow2\\
\mathit{WF}(\epsilon) & = & \top\\
\mathit{WF}(a) & = & \top\\
\mathit{WF}(A) & = & \mathit{WF}(R(A))\\
\mathit{WF}(e_{1}e_{2}) & = & \mathit{WF}(e_{1})\wedge(\nu(e_{1})\implies\mathit{WF}(e_{2}))\\
\mathit{WF}(e_{1}/e_{2}) & = & \mathit{WF}(e_{1})\wedge\mathit{WF}(e_{2})\\
\mathit{WF}(e^{\star}) & = & \mathit{WF}(e)\wedge\neg\nu(e)\\
\mathit{WF}(!e) & = & \mathit{WF}(e)\\
\mathit{WF}(\_) & = & \top\\
\mathit{WF}(\emptyset) & = & \top
\end{eqnarray*}
\end{minipage}

\caption{\label{fig:Nullability-and-WF}Nullability predicate $\nu(e)$ and
well-formedness predicate $\mathit{WF}(e)$.}
\end{figure}
In order to compute $\nu$ and $\mathit{WF}$, we must use Kleene's
fixed-point theorem, choosing $\bot$ as the least element and iterating
to stability, joining iterations with $\vee$. For example, the left-recursive
definition $X\leftarrow Xx/\epsilon$ is nullable, but not well-formed:
\begin{eqnarray*}
\nu(X)=\nu(R(X)) & = & \nu(Xx/\epsilon)\\
 & = & \nu(Xx)\vee\nu(\epsilon)\\
 & = & \nu(Xx)\vee\top=\top\\
\\
\mathit{WF}(X)=\mathit{WF}(R(X)) & = & \mathit{WF}(Xx/\epsilon)\\
 & = & \mathit{WF}(Xx)\wedge\mathit{WF}(\epsilon)\\
 & = & \mathit{WF}(X)\wedge(\nu(X)\implies\mathit{WF}(x))\\
 & = & \mathit{WF}(X)\wedge(\top\implies\top)\\
 & = & \mathit{WF}(X)=\bot\textrm{ (least element)}
\end{eqnarray*}
Its right-recursive variation $X'\leftarrow xX'/\epsilon$, however,
is both nullable and well-formed:
\begin{eqnarray*}
\mathit{WF}(X')=\mathit{WF}(R(X')) & = & \mathit{WF}(xX'/\epsilon)\\
 & = & \mathit{WF}(xX')\wedge\mathit{WF}(\epsilon)\\
 & = & \mathit{WF}(x)\wedge(\nu(x)\implies\mathit{WF}(X'))\\
 & = & \top\wedge(\bot\implies\bot)\\
 & = & \top\wedge\top=\top
\end{eqnarray*}
\begin{example}
Consider the PEG
\[
P\leftarrow\&c\;\;cc
\]
The input $cc$ is accepted if $\nu(D_{cc}P)$, that is, if $\nu(D_{c}(D_{c}P))$:
\begin{eqnarray*}
\nu(D_{c}(D_{c}P)) & = & \nu(D_{c}(D_{c}(\&c\quad cc)))\\
 & = & \nu(D_{c}(\&(\epsilon\quad\_^{*})\quad c))\\
 & \vdots\\
 & = & \nu(\&(\_^{*}\quad\_^{*})\quad\epsilon)\\
 & = & \nu(\&(\_^{*}\quad\_^{*}))\land\nu(\epsilon)\\
 & = & \nu(\_^{*}\quad\_^{*})\land\top\\
 & = & \nu(\_^{*})\land\nu(\_^{*})\\
 & = & \top
\end{eqnarray*}
\end{example}

\section{\label{sec:Implementing-derivatives}Implementing derivatives for
PEGs}

As of this writing, we have two implementations of the algorithms
described in this paper: one in JavaScript, and another in Racket
\cite{Flatt2010}. In both cases, as reported by Might et al. \cite{Might2011},
we found that careful attention to \emph{memoization}, \emph{laziness},
and \emph{fixed-point calculations} was required. In order to avoid
the accumulation of unimportant structure while computing derivatives
(which would make our implementations impractical) it was also necessary
to apply a set of grammar identities to simplify the expressions resulting
from functions $D$ and $\delta$.

Memoization is important for both $D$ and $\delta$ to avoid redundant
computation and space explosion. Furthermore, because PEGs as defined
here form a general directed graph, rather than a DAG or tree, our
implementations introduce laziness at nonterminals $A$ by memoizing
\emph{promises} in order to break cycles in the graph of the grammar.
The same approach was used by Bracha (via instances of \texttt{ForwardReferenceParser})
in his implementation of executable grammars in Newspeak \cite{Bracha2007}.

Our Racket implementation makes use of the memoization machinery for
Racket made available by Might et al. \cite{Might2011}, but we were
unable to use the associated implementation of iteration to fixed
point when implementing $\nu$, $\mathit{WF}$, and $\mathit{firsts}$.
Our definition of $\nu$ includes logical negation (in the clause
for negative lookahead) and so our fixed point algorithm must take
into account the join operator of the Boolean lattice being used,
on pain of non-termination in (ill-formed) cases such as computing
$\nu(A)$ when $A\leftarrow!A$. Might's implementation does not support
the specification of a join operator, instead simply updating variables
after each iteration. This fails not only for $\nu$ with ill-formed
PEGs but also for $\mathit{firsts}$ even with well-formed PEGs: the
first set for a nonterminal must be grown iteratively using a lattice
with least element $\emptyset$ and join operator $\cup$.

\begin{figure}[tb]
\[
\begin{array}[t]{lcclcclcl}
\&\epsilon\rightarrow\epsilon &  &  & !\epsilon\rightarrow\emptyset &  &  & \epsilon\;e\rightarrow e &  & \emptyset\ /\ e\rightarrow e\\
\&(\_^{\star})\rightarrow\epsilon &  &  & !(\_^{\star})\rightarrow\emptyset &  &  & e\;\epsilon\rightarrow e &  & e\ /\ \emptyset\rightarrow e\\
\&\emptyset\rightarrow\emptyset &  &  & !\emptyset\rightarrow\epsilon &  &  & \emptyset\;e\rightarrow\emptyset &  & e_{1}\ /\ !e_{1}\ e_{2}\rightarrow e_{1}\ /\ e_{2}\\
\&(e\;\_^{\star})\rightarrow\&e &  &  & !(e\;\_^{\star})\rightarrow!e &  &  & e\;\emptyset\rightarrow\emptyset &  & e_{1}\ /\ !e_{1}\rightarrow e_{1}\ /\ \epsilon\\
\&\&e\rightarrow\&e &  &  & !\&e\rightarrow!e &  &  & \_^{\star}\;\_^{\star}\rightarrow\_^{\star} &  & e\ /\ e\rightarrow e\\
\&!e\rightarrow!e &  &  & !!e\rightarrow\&e &  &  & !e_{1}\;!e_{1}\;e_{2}\rightarrow!e_{1}\;e_{2} &  & e_{1}\ e_{2}\ /\ e_{1}\ e_{3}\rightarrow e_{1}\ (e_{2}\ /\ e_{3})\\
 &  &  &  &  &  & \&e_{1}\;\&e_{1}\;e_{2}\rightarrow\&e_{1}\;e_{2} &  & e_{1}\;\_^{\star}\;/\;e_{2}\;\_^{\star}\rightarrow(e_{1}\;/\;e_{2})\;\_^{\star}
\end{array}
\]

\caption{\label{fig:Simplifications}Simplifications applied when constructing
PEG terms.}

\end{figure}
Implementations of the method of derivatives generally make heavy
use of identities to avoid constructing and processing needless structure.
For example, Brzozowski reports that without use of an equational
theory for REs, a given RE may have an unbounded number of distinct
derivatives, but with a simple set of identities, the number of derivatives
of an RE is finite \cite{Brzozowski1964}. Owens et al. enrich the
equational theory in order to more closely approach a minimal DFA
when using derivatives to compile REs \cite{Owens2009}. Finally,
Might et al. also employ identities in their \emph{compaction} of
CFGs during recognition, in between uses of $D$ \cite{Might2011}.
In our implementations, as in the implementation of Owens et al.,
most of the identity-based simplifications are placed in \emph{smart
constructors} for our PEG data type. Figure \ref{fig:Simplifications}
lists the identities respected by our implementations, including many
from previous work on derivatives as well as from Ford's original
work on PEGs \cite[§3.7]{Ford2004}. 

\section{\label{sec:Evaluation}Evaluation}

Our experience with using derivatives of PEGs to recognise and generate
sentences has led us to believe that our definition is correct, not
only with respect to the definition of derivatives originating with
Brzozowski \cite[definition 3.1]{Brzozowski1964},
\begin{quote}
Given a set $R$ of sequences and a finite sequence $s$, the \emph{derivative
of $R$ with respect to $s$} is denoted by $D_{s}R$ and is $D_{s}R=\{t\;|\;st\in R\}$
\end{quote}
but also with respect to Ford's original semantics for PEGs \cite{Ford2004}.
That is to say, we believe that the following conjectures hold, and
our experience has to date supported this belief.
\begin{conjecture}
For all well-formed expressions $e$ and terminals $a\in V_{T}$,
$\mathit{WF}(D_{a}e)$.
\end{conjecture}

Well-formedness is preserved by derivation.
\begin{conjecture}
For all well-formed expressions $e$ and strings $x\in V_{T}^{\star}$,
$(e\;\_^{\star},x)\Rightarrow\epsilon$ iff $\nu(D_{x}(e\;\_^{\star}))$.
\end{conjecture}

PEGs are defined by Ford in terms of matching a \emph{prefix} of their
inputs, but recognising via derivatives is defined in terms of the
\emph{entire} input. Consideration of $(e\;\_^{\star})$ ensures that
both definitions line up, either consuming the entirety of the input
or failing.
\begin{conjecture}
For all well-formed expressions $e$, terminals $a\in V_{T}$, and
strings $x,y\in V_{T}^{\star}$, $(e,axy)\Rightarrow y$ iff $(D_{a}e,xy)\Rightarrow y$.
\end{conjecture}

This connects Brzozowski's definition of derivatives with Ford's semantics.
If a PEG $e$ can accept input $axy$ yielding input suffix $y$,
then $e$ ``after the terminal $a$'', $D_{a}e$, should accept
input $xy$ yielding input suffix $y$, and vice versa.
\begin{conjecture}
For all well-formed expressions $e$ and strings $x\in V_{T}^{\star}$,
$\mathit{gen}(e)$ can produce $x$ iff $(e,x)\Rightarrow\epsilon$.
\end{conjecture}

Any string produced by our $\mathit{gen}$ algorithm is recognisable
by $e$, and any string recognisable by $e$ may be produced by $\mathit{gen}$.
\begin{conjecture}
For all well-formed expressions $e$, $\{a\;|\;ax\in L(e)\}\subseteq\mathit{firsts}(e)$.
\end{conjecture}

Our $\mathit{firsts}$ function (figure \ref{fig:peg-firsts-algorithm})
must be complete with respect to the actual language of its argument
in order for $\mathit{gen}$ to work correctly. It may, of course,
be larger than it needs to be, at the cost of some inefficiency in
$\mathit{gen}$.

\section{\label{sec:Related-Work}Related Work}

The notion of derivatives in grammatical formalisms is not new: Brzozowski
first introduced it in 1964, in the context of regular expressions
\cite{Brzozowski1964}. This work was later extended by Owens et al.
to generate efficient DFA-based recognisers for REs \cite{Owens2009},
and by Might et al. to generate parsers for CFGs. 50 years after Brzozowski's
original contribution, we extend the notion of derivatives to PEGs.
Independently, Moss has tackled this problem, taking quite a different
approach that augments the syntax of PEGs with the ongoing state of
a parse \cite{Moss2014}. By constrast, our approach does not require
any new syntax, and shows that PEGs are closed under derivative.

Applications of Brzozowski's derivatives are not limited to recognizing
and parsing. For example, they have also been used to compile Esterel
programs to automata \cite{Berry1999}, to generate DFAs from regular
expressions that are used to monitor program traces \cite{SenRosu2003},
and to prove the totality of parser combinators \cite{Danielsson2010}.
In this paper, we present a new application of derivatives to sentence
generation.

While there are many different techniques for generating sentences
from CFGs (e.g., random generation \cite{Seaman1974,Bird1983,Sirer1999,Forrester2000,Dreyfus2013},
exhaustive enumeration \cite{Mandl1985,Coppit2005,Majumdar2007,Godefroid2008},
and coverage-based generation \cite{Purdom1972,Lammel2006}), the
problem of generating sentences from PEGs has received very little
attention until now. To the best of our knowledge, Petit Parser \cite{Renggli2010}
is the only other system that can generate random sentences from a
PEG. However, Petit Parser's sentence generator ignores both positive
and negative lookahead expressions, which can lead to incorrect results.
This is a serious limitation because in addition to being the source
of much of the power of the PEG formalism\footnote{For example, the $a^{n}b^{n}c^{n}$ grammar discussed in section \ref{sec:Introduction}
makes essential use of lookahead \cite[§3.4]{Ford2004}.}, positive and negative lookahead expressions are frequently used
in practical grammars to (for example) correctly recognize keywords:
\begin{eqnarray*}
\mathit{While} & \leftarrow & 'while'\ !\mathit{Alnum}\\
\mathit{Return} & \leftarrow & 'return'\ !\mathit{Alnum}
\end{eqnarray*}
The $!\mathit{Alnum}$ expressions, if $R(\mathit{Alnum})$ matches
a single alphanumeric character, prevent the rules from matching identifiers
that contain their respective keyword as a prefix, e.g., $\mathit{while01}$
and $\mathit{returnedItems}$. By combining our definition of derivatives
for PEGs with our generic derivative-based sentence generation technique,
we are able to correctly generate sentences from any PEG, including
those that use positive and negative lookahead.

\section{\label{sec:FutureWork}Applications and Future work}

The ability to generate sentences from PEGs could give language designers
and implementers powerful new ways to write and debug grammars. For
example, we are currently working on an IDE for a PEG-based parser
generator that will show examples of sentences in the language defined,
even as the programmer edits the grammar\emph{.} Changes in the grammar
will be reflected immediately in the set of examples displayed, and
the programmer will be able to select some or all of the randomly-generated
examples to make them into unit tests that are run after each change.
That way, if a later edit results in one of those examples being rejected,
the programmer will know right away. We expect this kind of interface
will help programmers better understand their grammars, and enable
them to catch bugs earlier than with traditional tools.

Sentence generation could also be used to produce \emph{probabilistic}
\emph{proofs} of PEG identities, which can be difficult to reason
about. For example, in order to determine whether
\[
e_{1}\ e_{2}{^{*}}\ /\ e_{3}\ e_{2}{^{*}}\equiv(e_{1}\ /\ e_{3})\ e_{2}{^{*}}
\]
one might use a PEG grammar describing parsing expressions themselves
to generate random values for $e_{1}$, $e_{2}$, and $e_{3}$, then
substitute those values into both of the expressions above, and generate
a large number of sentences from each of them. If all sentences generated
by the left hand side of the potential identity are matched by the
expression on right hand side and vice-versa, there is a high probability
that they are equivalent.

While the initial motivation of the work presented in this paper was
sentence generation, our definition of derivative in section \ref{sec:peg-derivatives}
combines with the algorithm of section \ref{subsec:Recognising-with-derivatives}
to build a recogniser for PEGs. It would be interesting to go a step
further in this direction and adapt Might et al.'s technique for building
derivatives-based parser combinators for CFGs \cite{Might2011} to
PEGs.

Researchers have proposed extensions to the semantics of PEGs that
allow grammars to include left-recursive rules \cite{Warth2008,Medeiros2012}.
Without these extensions, an application of a rule that is left-recursive
will always result in infinite recursion, just as it would in a recursive-descent
parser. But the semantics of left recursion in PEGs remains a topic
of debate. For example, it is not at all clear whether a rule that
mixes left- and right-recursive applications should produce parse
trees that are left- or right-associative \cite{Tratt2010}. We believe
the notion of derivatives may help settle this debate. When parsing
with derivatives, choices are evaluated in parallel\textemdash i.e.,
the derivative of a choice expression ``makes progress'' on both
of its branches\textemdash so it may be possible to use the foundations
laid out in this paper to find a ``natural'' semantics for left-recursive
PEGs. %

Another extension of PEGs synergistic with our sentence generation
technique is OMeta \cite{Warth2007}. OMeta extends PEGs with an expressive
form of pattern matching that works on structured data (objects, arrays,
etc.) as well as strings. By extending our notion of derivative to
OMeta, we should be able to use grammars to generate random objects
for testing purposes, a la QuickCheck \cite{Claessen2000}.

PEG sentence generation has other applications in testing. For example,
it should now be possible to build PEG-based equivalents of popular
grammar-based testing tools such as CESE \cite{Majumdar2007} and
SAGE \cite{Godefroid2008}, both of which are based on CFGs. Our random
sentence generation technique could also be used to ensure (probabilistically)
that a compiler can handle all possible nestings / interactions among
the constructs that are available in a dynamic programming language.
(Doing the same for the compiler of a statically-typed language would
require generating well-typed programs, which is still an open problem.)

\textbf{}%

\section{\label{sec:Conclusion}Conclusion}

In this paper, we have adapted Brzozowski's notion of derivatives
to PEGs, and introduced a new sentence generation technique based
on derivatives that is applicable to any grammatical formalism for
which the derivative can be defined, which now includes PEGs. We have
also outlined some ways in which sentence generation can improve tool
support for language designers and implementers who use PEGs to define
the syntax of their languages.

\end{comment}
\end{thebibliography}

\end{document}